\documentstyle[12pt,draft]{article}

\newcommand{\beq}{\begin{equation}}
\newcommand{\eeq}{\end{equation}}
\newcommand{\ber}{\begin{eqnarray}}
\newcommand{\eer}{\end{eqnarray}}

\newcommand{\teq}{\hskip-.1in&=&\hskip-.1in}
\newcommand{\tab}{\hskip-.265in&&\hskip-.1in}
\newcommand{\tsim}{\hskip-.1in&\sim&\hskip-.1in}

\newcommand{\vev}[1]{{\left< {#1} \right>}}
\newcommand{\inv}[1]{{1\over{#1}}}

\newcommand{\Tr}{{\rm Tr}}
\newcommand{\cN}{{\cal N}}
\newcommand{\cA}{{\cal A}}
\newcommand{\cD}{{\cal D}}
\newcommand{\cL}{{\cal L}}
\newcommand{\cP}{{\cal P}}
\newcommand{\cS}{{\cal S}}
\newcommand{\half}{{1\over 2}}
\newcommand{\quart}{{1\over 4}}

\newcommand{\mysection}[1]{\setcounter{equation}{0}\section{#1}}

\begin{document}

\begin{titlepage}

\rightline{NSF-ITP-99-99}
\rightline{hep-th/9908113}

\begin{center}

\vskip 1.5 cm
{\Large \bf A New Type of Loop Equations}

\vskip 1 cm
{\large Nadav Drukker}\\

\vskip .5cm
Department of Physics, Princeton University,\\
Princeton, NJ 08544

\smallskip
{\it and}
\smallskip

Institute for Theoretical Physics,
University of California,\\
Santa Barbara, CA 93106

\bigskip
{\tt drukker@itp.ucsb.edu}

\end{center}

\vskip 0.5 cm
\begin{abstract}
We derive a new form of loop equations for light-like Wilson loops.
In bosonic theories those loop equations close only for straight 
light-like Wilson lines.  In the case of $\cN=1$ in ten dimensions 
they close for any light-like Wilson loop.  Upon 
dimensional reduction to $\cN=4$ SYM in four dimensions, these loops 
become exactly the chiral loops which can be evaluated 
semiclassically, in the strong coupling limit, by a minimal surface in 
anti de-Sitter space.  We show that the $AdS$ calculation satisfies 
those loop equations.  We also find a new fermionic 
loop equation derived from the gauge theory fermionic equation of 
motion.
\end{abstract}
\end{titlepage}

\mysection{Introduction}

Loop equations were proposed \cite{Makeenko:1979pb} in an attempt 
to understand the string theory that governs confinement in QCD. 
The loop equations are dynamical equations for the Wilson loop 
\cite{Wilson:1974sk}, which in the string picture is the boundary 
of an open string. The equation is given by a second order 
differential operator which annihilates smooth loops, and 
gives a non-trivial result for self-intersecting loops. They can 
be proven by a formal derivation, and therefore should be solved 
by any (regularized, but not renormalized) calculation of 
the Wilson loop. Indeed, they are solved by perturbation theory 
\cite{Brandt:1982gz} and by lattice QCD. In the case of two 
dimensional pure gauge, which is soluble, this solution also 
solves the loop equation\cite{Kazakov:1980zi}.

An important feature of loop equations is that they are correct 
at all values of the coupling constant. That is the basis for the 
hope of understanding confinement through them. In particular, 
they constitute a test for any proposed definition of the theory 
at strong coupling.

The famous Maldacena conjecture 
\cite{Maldacena:1997re,Gubser:1998bc,Witten:1998qj}\footnote{for 
a recent review see \cite{Aharony:1999ti}} 
states that $\cN=4$ supersymmetric $U(N)$ gauge theory in four 
dimensions is dual to type IIB string theory on a background 
geometry $AdS_5\times S^5$ (with certain 5-form flux). According 
to the conjecture \cite{Maldacena:1998im,Rey:1998ik} the 
expectation value of a Wilson loop is given at large $N$ and 
strong coupling by a minimal surface calculation. This is exactly 
such a non-perturbative definition of a gauge theory, and it is 
natural to ask if this ansatz solves the loop equation.

This question was addressed recently \cite{Drukker:1999zq}, 
and the conclusion was that to the extent that the equations 
could be checked they are solved by string theory. 
The main difficulty in applying the loop equations to the 
$AdS$ calculation was that the $AdS$ ansatz is simple only for 
a restricted family of loops. If the loop is smooth its 
expectation value is given by a minimal surface only if it
satisfies a local BPS condition that the coupling to the scalars 
is equal to the coupling to the gauge fields. If one writes 
the Wilson loop (in Lorentzian space) as
\beq
W={1\over N}\Tr\,\cP\, \exp\left(
i\oint\left(A_\mu\dot x^\mu+\Phi_i\dot y^i\right)ds\right),
\eeq
where $A_\mu$ are the gauge fields and $\Phi_i$ the scalars, 
the condition is $|\dot x|=|\dot y|$.

The regular definition of the loop Laplacian requires 
differentiation in all directions, and so breaks this constraint. 
Therefore in that paper we had to guess how to extend the ansatz 
also to loops which break the constraint slightly. There is a 
bigger problem for loops with cusps or intersections, then the 
constraint is broken by a large amount and therefore the loop 
equations could not be checked.

In this paper we find a new form for the loop equations 
which does not require differentiation away from the 
constraint. One differentiates only in eight independent 
bosonic and eight fermionic directions which commute with the 
constraint. Those equations also behave better under 
supersymmetry, and in fact the loop Laplacian is the 
anticommutator of two fermionic operators. One of the 
fermionic operators generates another loop equation derived 
from the gauginos' equations of motion.

Those equations can be derived also in non-supersymmetric gauge 
theories, but only for a very restricted family of loops. The 
Wilson loop has to be a straight light-like line.

In the next section we review the regular form of the loop 
equations (for non-supersymmetric theories).

The derivation of the light-cone loop equations is presented in section 
\ref{derivation}.  The proofs of some of the equations presented in 
that section are given in the appendix, because they are rather 
horrible.

We conclude with a discussion on how those loop equations can be 
applied to the Wilson loops that are calculated by minimal surfaces 
in $AdS$. The equations are satisfied by smooth loops with no self 
intersections. They cannot be checked when the loops have cusps or 
intersections, since those loops cannot be calculated by a minimal 
surface when the constraint is satisfied. We also speculate about 
other applications for this formalism, and present some suggestions 
for future work.

\mysection{Short Review of Loop Equations}
\label{review}

Let us first review the regular form of the loop equation in pure 
Yang-Mills, as the new form will use many of the same techniques. 
For more complete reviews see \cite{Migdal:1984gj, Polyakov:1987ez}.

The action of pure gauge theory in any number of dimensions
is\footnote{The complete action contains a gauge fixing term and
ghosts. Those appear also in the equations of motion, but
can be dropped by a Ward identity \cite{Brandt:1982gz}.}
\beq
\cS=-{1\over4g_{YM}^2}\int dx\,\Tr F_{\mu\nu}F^{\mu\nu},
\eeq
and the Wilson loop is given by
\beq
W={1\over N}\Tr\,\cP\, \exp\left(i\oint A_\mu dx^\mu\right),
\eeq
where the integral is over a path parametrized by~$x^\mu$.

The first functional derivative of the loop is
\beq
{\delta\over\delta x^\mu(s)}W
=-i\dot x^\nu(s) F_{\nu\mu}^a
\Tr\cP\left[T^a(s) \exp\left(i\oint A_\mu dx^\mu\right)\right],
\eeq
$F_{\mu\nu}$ is the field strength and $T^a(s)$ is a generator 
of the gauge group inserted at the point $s$ along the loop.

The second derivative can insert another $F_{\mu\nu}$ into the 
loop, but also has a contact term
\ber
{\delta^2\over\delta x^\mu(s)\delta x_\mu(s')}W
\teq
-\dot x^\nu(s) F_{\nu\mu}^a\,\dot x^\rho(s') F_\rho^{~\mu b}\,
\Tr\cP\left[T^a(s)T^b(s')\exp\left(i\oint A_\mu dx^\mu\right)\right]
\nonumber\\
\tab
-i\dot x^\nu(s) D^\mu F_{\nu\mu}^a \delta(s-s')
\Tr\cP\left[T^a(s) \exp\left(i\oint A_\mu dx^\mu\right)\right].
\nonumber\\
\label{second-deriv}
\eer
To write the loop equations we want to retain only the contact 
term. Therefore we integrate $s'$ over an interval of vanishing 
width around $s$, picking out only the delta function term.
The loop Laplacian is defined as
\beq
\hat L=\lim_{\eta\to 0}
\oint ds \int_{s-\eta}^{s+\eta} ds'
{\delta^2\over\delta x^\mu(s')\delta x_\mu(s)}.
\label{boson-deriv}
\eeq
From (\ref{second-deriv}) we get
\beq
\hat L\,\vev{W}=-i\oint ds\,\dot x^\mu \vev{(D^\nu F_{\mu\nu})^a(s)
\,{1\over N}\Tr\cP
\left[T^a(s)\exp\left(i\oint A_\mu dx^\mu\right)\right]}.
\label{regular-L}
\eeq
So we see that $\hat L$ inserts the variation of the action 
$\dot x^\mu D^\nu F_{\mu\nu}$ into the loop.
This insertion into the loop would
be zero if we use the classical equation of motion,
but quantum corrections produce contact terms.  To see
that, one can write the equations of motion as the functional
derivative of the action $\cS$ and use the Schwinger-Dyson
equations,
$i.e.$ integration by parts in the functional integral,
\ber
\hat L\,\vev{W}
\teq g_{YM}^2\int\cD A\oint ds\,{1\over N}\Tr\cP
\left[T^a(s)\exp\left(i\oint A_\mu dx^\mu\right)\right]
\dot x^\mu(s){\delta e^{i\cS} \over\delta A^{\mu a}(x(s))}
\nonumber\\
\teq -g_{YM}^2
\vev{\oint ds\,\dot x^\mu(s){\delta\over\delta A^{\mu a}(x(s))}
{1\over N}\Tr\cP\left[T^a(s)\exp\left(i\oint A_\mu dx^\mu\right)\right]}.
\nonumber\\
\eer
The functional derivative $\delta/\delta A_\mu(x(s))$
in this equation is formally evaluated as
\ber
\hat L\,\vev{W}
\teq-i{\lambda\over N^2}\oint ds\oint ds'\, \delta(x^\mu(s')-x^\mu(s))
\dot x_\mu(s)\dot x^\mu(s')
\nonumber\\ 
\tab\hskip.7in\times
\vev{\Tr\,\cP\left[ T^a(s)T^a(s')
\exp\left(i\oint A_\mu dx^\mu\right)\right]}.
\eer
We then use the relation between the generators of $SU(N)$,
\beq
T^a_{nm}T^a_{kl}
=\delta_{nk}\delta_{ml}-{\delta_{nm}\delta_{kl}\over N}.
\eeq
Ignoring the $1/N$ term, the trace is broken into two.
This gives the
correlation function of two loops. In the large $N$ limit,
the correlator factorizes and we obtain,
\beq
\hat L\,\vev{W}
=-i\lambda\oint ds\oint ds'\, \delta(x^\mu(s')-x^\mu(s))
\dot x_\mu(s)\dot x^\mu(s') \vev{W_{ss'}}\vev{W_{s's}}.
\label{boson-eqn}
\eeq
Here $W_{ss'}$ is a Wilson-loop that start at~$s$ and goes to~$s'$
and~$W_{s's}$ goes from~$s'$ to~$s$.  They are closed due to
the delta function.

According to equation (\ref{boson-eqn}) $\hat{L} \langle W \rangle$
should receive contributions from self-in\-ter\-sec\-tions of 
the loop. Every intersection point contributes a term proportional 
to the dot product of the two tangent vectors. Likewise, 
at a cusp we expect to get a term proportional to the product of 
the right and left tangent vectors. 
It seems like we should also count the trivial
case of~$s=s'$, when $W_{ss'}=1$ and $W_{s's}=W$.
In most of the literature on the loop equation, this trivial
self-intersection is ignored. In any case, 
it can be taken care of by multiplicative renormalization 
of the loop operator. For a supersymmetric loop, there 
should be no renormalization and the leading contribution from 
the trivial self-intersection cancels, because $\dot x^2=0$.

\mysection{Light-Cone Loop Equations}
\label{derivation}

After presenting the regular loop equations, we derive the 
light-cone loop equations, utilizing constrained derivatives. 
First we derive them for bosonic theories (where they usually 
don't give closed loop equations), and then for the 
supersymmetric case. Somewhat similar ideas were used in the 
context of the type IIB matrix model in 
\cite{Fukuma:1997en,Hamada:1997dt}.

\subsection{Bosonic Case}

The bosonic light-cone loop equations can be derived in any dimension.  
But since we will eventually turn to the $\cN=4$ theory in four 
dimensions, we will do it for the bosonic sector of that theory.  For 
other cases there are changes in numerical factors.

The theory has four gauge fields and six scalars. We work in 
ten dimensional notations, where they are all denoted by 
$A_\mu$ ($\mu=0\ldots9$). Because Euclidean supersymmetry is 
a bit tricky, we work in Lorentzian signature $(-,+,\cdots,+)$.

Our goal is to write loop equations for loops satisfying the 
on-shell condition $m^2=\dot x_\mu \dot x^\mu=0$.  We use light-cone 
coordinates $\dot x_\pm=\left(\pm\dot x_0+\dot x_1\right)/\sqrt2$ 
so the constraint is $2\dot x_-\dot x_+ + \dot x_i\dot x_i 
= 0$.  A variation that preserves this constraint will be of 
the form 
$\dot x_-\delta\dot x_++\dot x_+\delta\dot x_- +\dot 
x_i\delta\dot x_i=0$. 
We take $\delta x_i$ to be the independent variations; then 
one can use reparametrization invariance to set $\dot x_-$ 
to a constant and $\delta\dot x_-=0$.  Then solve
\beq
\delta x_+
= -\int_0^s ds' {\dot x_i\delta\dot x_i \over \dot x_-}
= -{\dot x_i \over \dot x_-}\delta x_i
+\int_0^s ds' {\ddot x_i\delta x_i \over \dot x_-}.
\label{xplus}
\eeq
so
\beq
{\delta x_+(s)\over\delta x_i(s')}
= -{\dot x_i \over \dot x_-}\delta(s-s')
+\hbox{less singular}
\eeq
The contribution from the second term in (\ref{xplus}) is less 
singular, because of the integral. Since we are interested only 
in the most singular contact term it can be ignored.

Therefore we can define a differential operator that commutes with 
the constraint
\beq
{\delta\over\delta x_i}
\sim
{\partial\over\partial x_i}
-{\dot x_i\over\dot x_-}{\partial\over\partial x_+}.
\label{cvar}
\eeq
The derivatives on the right hand side are unconstrained functional 
derivatives.

We define the bosonic light-cone loop Laplacian
\beq
\hat L^{lc}_B
=\lim_{\eta\to 0}\oint ds\int_{s-\eta}^{s+\eta} ds'
{\delta^2\over\delta x_i(s)\delta x_i(s')}.
\eeq
Naively one would expect that to be\footnote{
To reduce clutter, we drop the integration signs in this equation, 
but on the right hand side we still mean that only the contact 
term is picked out and the expression is integrated around the loop. 
The same will be done in many of the equations that follow, where 
some obvious parts of the equations are omitted.}
\ber
\hat L^{lc}_B
\teq
\left({\partial\over\partial x_i}
-{\dot x_i\over\dot x_-}{\partial\over\partial x_+}
\right)
\left({\partial\over\partial x_i}
-{\dot x_i\over\dot x_-}{\partial\over\partial x_+}
\right)
\nonumber\\
\teq
{\partial^2\over\partial x_i\partial x_i}
-{\dot x_i\over\dot x_-}\left(
{\partial^2\over\partial x_i\partial x_+}
+{\partial^2\over\partial x_+\partial x_i}\right)
+{\dot x_i^2\over\dot x_-^2}
{\partial^2\over\partial x_+\partial x_+}
+{8\over\dot x_-}{d\over ds}
\left({\partial\over\partial x_+}\right),
\nonumber\\
\eer
where the last term comes from the cross term, 
$\partial/\partial x_i$ acting on $\dot x_i$ (with a factor 
of 8 from tracing over the index $i$). But a more careful 
calculation shows that the second variation of $x_+$ is
\beq
\delta^2 x_+
\sim-{1\over\dot x_-}\delta x_i\delta\dot x_i
=-{1\over2\dot x_-}{d\over ds}\left(\delta x_i\right)^2.
\label{delta2x+}
\eeq
The factor of a half on the right hand side means that the cross 
term should be decreased by a half.

Now we are ready to evaluate the action of this operator on the 
Wilson loop.
\ber
\hat L^{lc}_B
\teq
{\partial^2\over\partial x_i\partial x_i}
-{\dot x_i\over\dot x_-}\left(
{\partial^2\over\partial x_i\partial x_+}
+{\partial^2\over\partial x_+\partial x_i}\right)
+{\dot x_i^2\over\dot x_-^2}{\partial^2\over\partial x_+\partial x_+}
+{4\over\dot x_-}{d\over ds}
\left({\partial\over\partial x_+}\right)
\nonumber\\
\teq
-i\dot x^\mu D_iF_{\mu i}
+i{\dot x_i\dot x^\mu\over \dot x_-}
\left(D_-F_{\mu i}+D_iF_{\mu -}\right)
-i{\dot x_i^2\dot x^\mu\over \dot x_-^2}D_-F_{\mu -}
-{4i\over\dot x_-}{d\over ds}(\dot x^\mu F_{\mu -})
\nonumber\\
\teq
-i\dot x^\mu D^\nu F_{\mu\nu}
+i{\dot x^\mu \dot x^\nu\over\dot x_-} D_\nu F_{\mu -}
-{4i\over\dot x_-}{d\over ds}(\dot x^\mu F_{\mu -}).
\label{rough-bosons}
\eer
The last term is a total derivative, so for a closed (topologically 
trivial) loop it integrates to zero.

For a straight line $\ddot x=0$, so the second term is the same 
as the third and is also a total derivative. Ignoring the total 
derivatives we find
\beq
\hat L^{lc}_B\vev{W}
=
-{i\over N}\oint ds\,\dot x^\mu(s) \vev{D^\nu F_{\mu\nu}^a(x(s))\Tr\cP
\left[T^a(s)\exp\left(i\oint A_\mu \dot x^\mu ds\right)\right]}.
\eeq
This is the same as what the unconstrained loop Laplacian gives. 
Going through the regular procedure 
(\ref{regular-L})--(\ref{boson-eqn}) 
we get
\ber
\hat L^{lc}_B\vev{W}
\teq
-i{g_{YM}^2\over N}\oint ds\oint ds'\,\dot x^\mu(s)\dot x_\mu(s')
\delta(x(s)-x(s'))
\vev{W_1}\vev{W_2}
\nonumber\\
\tsim
-i{\lambda\over\epsilon^3}\oint ds\,\dot x^\mu\dot x_\mu
\vev{W}
=
0+O\left({\lambda\over\epsilon}\right),
\eer
plus nonzero contributions at intersections. 

The regular loop equations can be defined locally on the loop, 
not integrated around it. The constrained loop equations work 
only when integrating along the loop.

If the Wilson loop is not a straight line, the constrained 
variation does not give a closed loop equation. But the 
extra term 
$i{\dot x^\nu\dot x^\mu\over\dot x_-}D_\nu F_{\mu -}$ 
is canceled in the supersymmetric case.

\subsection{Supersymmetric Wilson Loop}

The $\cN=4$ gauge theory in four dimensions has, in addition 
to the gauge fields and scalars, also gauginos $\Psi$. The 
Wilson loop could couple also to them, or in other words, 
the bosonic loop we considered so far belongs to a 
supermultiplet, whose other members have also fermionic 
parameters. 
Still working in ten dimensional notations, in the light-cone 
$\Psi$ breaks into 
$\Psi_1$ in the $8_s$ of $SO(8)$ and $\Psi_2$ in the $8_c$ 
($\Gamma_-\Psi_1=\Gamma_+\Psi_2=0$).

The supersymmetry generators of the gauge theory $Q$ also 
decompose in the light-cone. $Q_1$ (after rescaling) 
acts by
\ber
2^\quart\sqrt{\dot x_-}[Q_{1a},A_\mu]
\teq
{i\over 2}\delta^i_\mu\gamma_{ia\dot a}\Psi_{2\dot a}
+{i\over\sqrt{2}}\delta_\mu^+\Psi_{1a},
\nonumber\\
2^\quart\sqrt{\dot x_-}\{Q_{1a},\Psi_{1b}\}
\teq
-{1\over4}\gamma_{ijba}F_{ij}+\half\delta_{ab}F_{-+},
\nonumber\\
2^\quart\sqrt{\dot x_-}\{Q_{1a},\Psi_{2\dot a}\}
\teq
{1\over\sqrt{2}}\gamma_{ia\dot a}F_{-i},
\label{Q1}
\eer
and $Q_2$ by
\ber
2^\quart\sqrt{\dot x_-}[Q_{1a},A_\mu]
\teq
{i\over 2}\delta^i_\mu\gamma^T_{i\dot aa}\Psi_{1a}
-{i\over\sqrt{2}}\delta_\mu^-\Psi_{2\dot a},
\nonumber\\
2^\quart\sqrt{\dot x_-}\{Q_{2\dot a},\Psi_{1a}\}
\teq-{1\over\sqrt{2}}\gamma_{i\dot aa}F_{+i},
\nonumber\\
2^\quart\sqrt{\dot x_-}\{Q_{2\dot a},\Psi_{2\dot b}\}
\teq-{1\over4}\gamma_{ij\dot b\dot a}F_{ij}
-\half\delta_{\dot a\dot b}F_{-+}.
\label{Q2}
\eer

As was shown in \cite{Drukker:1999zq}, the loops satisfying the 
masslessness condition $\dot x^2=0$ are locally BPS, and depend 
on only eight fermionic parameters. Therefore, to
write the supersymmetric Wilson loop we introduce a fermion 
$\zeta_a$ in the $8_s$. The supersymmetric Wilson loop is
\ber
W\teq
\inv{N}\Tr\,\cP\left[
\exp\left(\oint\zeta(s)(Q_1+\half\widetilde Q_1)ds\right)
\exp\left(i\oint A_\mu \dot x^\mu\,ds\right)
\right.\nonumber\\
\tab\left.\hskip1.5in
\times
\exp\left(-\oint\zeta(s)(Q_1+\half\widetilde Q_1)ds\right)
\right].
\eer
$\widetilde Q_1$ (and $\widetilde Q_2$) are other SUSY generators acting 
on $\dot x^\mu$ giving $\dot\zeta$ by
\ber
\sqrt{2}\dot x_-[\widetilde Q_{1a},\dot x_\mu]
\teq-{i\over\sqrt{2}}\delta_\mu^+\dot\zeta_a,
\nonumber\\
\sqrt{2}\dot x_-[\widetilde Q_{2\dot a},\dot x_\mu]
\teq-{i\over2}\delta_\mu^i\gamma_{i\dot aa}\dot\zeta_a.
\label{qhat}
\eer
There are a few justifications for including the action of 
$\widetilde Q_1$. For one we were unable to write loop equations 
without it. But also, regarding the Wilson loop as the first 
quantized action for a w-boson of a broken $SU(N+1)$ 
gauge group, we should use the supersymmetry generators 
of that group, which also act on the $\dot x$'s.

It is useful to write the supersymmetric loop as
\beq
W=
\inv{N}\Tr\,\cP\left[
\exp\left(i\oint \cA_\mu \dot x^\mu\,ds\right)
\right],
\label{cA}
\eeq
where in the exponent we have
\ber
i\cA_\mu\dot x^\mu
\teq
\exp\left(\zeta(Q_1+\half\widetilde Q_1)\right)
i A_\mu \dot x^\mu
\exp\left(-\zeta(Q_1+\half\widetilde Q_1)\right)
\nonumber\\
\teq
iA_\mu \dot x^\mu
-{\sqrt{\dot x_-}\over 2^{5\over 4}}
\left(\sqrt{2}\zeta\Psi_1
+{\dot x_i\over\dot x_-}\zeta\gamma_i\Psi_2\right)
+{1\over 4\dot x_-}A_-\zeta\dot\zeta
\nonumber\\
\tab
+{3\over 16\dot x_-}
\dot x_{[i} F_{j-]}\zeta\gamma_{ij}\zeta
+{i\over 2^\quart48\sqrt{\dot x_-}}
\zeta\gamma_{ij}\zeta
\left(D_i-{\dot x_i\over\dot x_-}D_-\right)\zeta\gamma_j\Psi_2
+\ldots
\nonumber\\
\label{exponent}
\eer
To simplify the equations we used the anti-symmetric symbol $[\ldots]$ 
(normalized by $1/6$).

The action of $\widetilde Q_1$ gives 
${1\over4\dot x_-}A_-\zeta\dot\zeta$. It is natural to absorb 
$\zeta\dot\zeta$ into $\dot x_+$, since both couple to $A_-$. 
This modifies the on-shell condition to
$2\dot x_-\dot x_++\dot x_i\dot x_i+{i\over 2}\zeta\dot\zeta=0$. 
This way, we can define a 
constrained and unconstrained fermionic variation, as in the 
bosonic case. It is
\beq
{\delta\over\delta \zeta}
\sim
{\partial\over\partial \zeta}
+{i\zeta\over4\dot x_-}{\partial\over\partial x_+},
\label{fcvar}
\eeq
which we use below.

\subsection{Supersymmetric Loop Equation}

The constrained supersymmetric loop Laplacian is
\beq
\hat L^{lc}
=
\lim_{\eta\to0}\oint ds\int_{s-\eta}^{s+\eta}ds' \left(
{\delta^2\over\delta x_i(s)\delta x_i(s')}
-i
{\delta\over\delta\zeta(s)}
\mathop{d\over ds}^{\leftrightarrow}
{\delta\over\delta\zeta(s')}
\right).
\label{hat-L}
\eeq

The $\zeta=0$ part of the bosonic variation is still
\beq
\hat L^{lc}_{B[0]}
=
-i\dot x^\mu D^\nu F_{\mu\nu}
+i{\dot x^\mu \dot x^\nu\over\dot x_-}D_\nu F_{\mu -},
\eeq
plus the total derivative.

To this we want to add the fermionic piece. 
When $d/ds$ acts to the right the $\zeta=0$ piece 
is
\ber
{\delta\over\delta\zeta}
{d\over ds}
{\delta\over\delta\zeta}_{[0]}
\tsim
\left({\partial\over\partial\zeta}
+{i\zeta\over 4\dot x_-}{\partial\over\partial x_+}
\right)
{d\over ds}
\left({\partial\over\partial\zeta}
+{i\zeta\over 4\dot x_-}{\partial\over\partial x_+}
\right)
\nonumber\\
\tsim
-{1\over 4\sqrt{2}\dot x_-}
\left\{\left(
\sqrt{2}\dot x_-\Psi_1+\dot x_i\gamma_i\Psi_2\right),
\left(\sqrt{2}\dot x_-\Psi_1+\dot x_j\gamma_j\Psi_2
\right)\right\}
\nonumber\\
\tab
+{2\dot x^\nu\dot x^\mu\over \dot x_-}D_\nu F_{\mu -}
\nonumber\\
\tsim
-\quart\dot x^\mu
\left\{\bar\Psi,\Gamma_\mu\Psi\right\}
+{2\dot x^\nu\dot x^\mu\over \dot x_-}D_\nu F_{\mu -}.
\label{rough-fermions}
\eer
As in (\ref{delta2x+}) there is a correction to the second 
derivative. Again it is equal to minus half the cross term, 
decreasing it to 
${\dot x^\nu\dot x^\mu\over\dot x_-}D_\nu F_{\mu -}$.

When $d/ds$ acts to the left you find only the first term in 
(\ref{rough-fermions}).  Together this is
\beq
\hat L^{lc}_{F[0]}
=
-\half\dot x^\mu
\left\{\bar\Psi,\Gamma_\mu\Psi\right\}
+{\dot x^\nu\dot x^\mu\over\dot x_-}D_\nu F_{\mu -}.
\eeq

Combining this with the bosonic piece gives
\ber
\hat L^{lc}_{[0]}\vev W
\teq
-{i\over N}\oint ds\,\dot x^\mu \left<
\left(
D^\nu F_{\mu\nu}^a
-\half\left\{\bar\Psi,\Gamma_\mu\Psi\right\}^a
\right)
\right.\nonumber\\
\tab\hskip.7in\left.
\times
\Tr\,\cP\left[T^a(s)
\exp\left(i\oint \cA_\mu \dot x^\mu\, ds\right)
\right]\right>.
\label{boson+fermion}
\eer
This is the equation of motion for the supersymmetric gauge 
theory, which allows us to complete the loop equations. 
By the usual manipulations this is equal to
\ber
\hat L^{lc}_{[0]}\vev W
\teq
-i\lambda
\oint ds\oint ds' \, \dot x^\mu(s) \dot x_\mu(s')
\delta^4(x(s)-x(s'))
\vev{W_1}\vev{W_2}.
\label{almost-right}
\eer
When $\delta/\delta A^\mu$ acts on the bosonic loop it brings down 
$i\dot x_\mu$, but acting on the supersymmetric loop (\ref{cA}) 
will bring down more terms of higher order in $\zeta$. It is easy 
to see, though, that all those terms are zero when multiplied by 
$\dot x^\mu$.

In the above we considered only the $\zeta=0$ part of the action of 
$\hat L^{lc}$ on the loop. When $\zeta=0$ the on-shell constraint is 
$\dot x^2=0$, so the right hand side vanishes for a smooth loop. But 
when $\zeta\neq0$, as we mentioned before, the constraint is modified 
to $\dot x_\mu\dot x^\mu+{i\over2}\zeta\dot\zeta=0$, so we expect to get 
this extra term at higher orders in $\zeta$. Indeed  calculating 
the term linear in $\zeta$ one finds many terms, most of them 
hopefully cancel, but some terms that remain are
\beq
\hat L^{lc}_{[1]}
=-{1\over2^\quart\sqrt{\dot x_-}}
\dot\zeta\left(\gamma_iD_i\Psi_2+\sqrt2D_-\Psi_1\right)
=-{1\over2^\quart\sqrt{\dot x_-}}
\dot\zeta\left(\Gamma^0\Gamma^\mu D_\mu\Psi\right)_1,
\label{hat-L1}
\eeq
where the right hand side is half of the components of the fermionic 
equations of motion
\beq
{\delta\cS\over\delta\bar\Psi}
=-{i\over g_{ym}^2}\Gamma_\mu D^\mu\Psi.
\label{fermionic-eom}
\eeq
We can use this to write a Schwinger-Dyson equation, or a fermionic 
loop equation. This will insert into the loop a derivative with 
respect to $\Psi_1$
\ber
\hat L^{lc}_{[1]}\vev W
\teq
{1\over N}
\oint ds{-\dot\zeta(s)\over2^\quart\sqrt{\dot x_-}} \vev{
{\delta\over\delta\Psi_1^a(x(s))}
\Tr\,\cP\left[T^a(s)
\exp\left(i\oint \cA_\mu \dot x^\mu\, ds\right)
\right]}
\nonumber\\
\teq
-\lambda\oint ds\oint ds'
\half\dot\zeta(s)\zeta(s')
\delta^4(x(s)-x(s'))\vev{W_1}\vev{W_2}.
\eer
Combined with (\ref{almost-right}) we get
\ber
\hat L^{lc}\vev W
\teq
-i\lambda\oint ds\oint ds'
\left(\dot x^\mu(s) \dot x_\mu(s')
-{i\over2}\dot\zeta(s)\zeta(s')\right)
\nonumber\\
\tab\hskip1.5in\times
\delta^4(x(s)-x(s'))\vev{W_1}\vev{W_2},
\eer
and the contribution from a smooth point is proportional to 
$\dot x_\mu\dot x^\mu-{i\over2}\dot\zeta\zeta$, which is indeed zero. 
If the loop has cusps or intersections there will be a contribution 
from all those points proportional to $i\lambda/\epsilon^2$ where 
$\epsilon$ is a UV cutoff and a function of the angle between the 
two tangent vectors.

There might be other terms at higher orders in $\zeta$, but we should
hope they are also zero. It would be 
nice to find a manifestly supersymmetric derivation of those 
equations, or another way to check them to all orders in~$\zeta$.

\subsection{Fermionic Identities}

The SUSY generators of the gauge theory (\ref{Q1}) and (\ref{Q2}) 
have the algebra
\ber
\left\{Q_{1a},Q_{1b}\right\}
\teq{i\over 2\dot x_-}\delta_{ab}D_-,
\nonumber\\
\left\{Q_{1a},Q_{2\dot b}\right\}
\teq{i\over 2\sqrt{2}\dot x_-}\gamma_{ia\dot b}D_i,
\nonumber\\
\left\{Q_{2\dot a},Q_{2\dot b}\right\}
\teq{-i\over 2\dot x_-}\delta_{\dot a\dot b}D_+.
\label{susy-algebra}
\eer

This algebra is realized on loops by the operators
\ber
q_1\teq\oint ds\left({\partial\over\partial\zeta}
-{i\zeta\over4\dot x_-}{\partial\over\partial x_+}\right),
\\
q_2\teq{1\over\sqrt2\dot x_-}\oint ds
\left(\dot x_i\gamma_i q_1
-{i\over 2}{\delta\over\delta x_i}\gamma_i\zeta\right)
\nonumber\\
\teq{1\over\sqrt2\dot x_-}\oint ds
\left(\dot x_i\gamma_i{\partial\over\partial\zeta}
-{i\over 2}{\partial\over\partial x_i}\gamma_i\zeta
+{i\over4\dot x_-}\dot x_i\gamma_i\zeta{\partial\over\partial x_+}\right).
\label{susy-generators}
\eer
We prove in the appendix that when acting on a loop $q_1=Q_1$ 
and $q_2=Q_2$ by an explicit calculation for the first few 
orders in $\zeta$. Part of the calculation is simply based on the 
construction of the loops in terms of $\zeta Q_1$, but at higher 
orders in $\zeta$ the calculation becomes non-trivial involving 
Fierz identities and loop equations. One can also check the 
commutators of those operators, it is easy to see that
\ber
\left\{q_{1a},q_{1b}\right\}
\teq{-i\over 2\dot x_-}
\delta_{ab}\oint ds\, {\partial\over\partial x_+},
\nonumber\\
\left\{q_{1a},q_{2\dot b}\right\}
\teq{-i\over 2\sqrt{2}\dot x_-}\gamma_{ia\dot b}
\oint ds\,{\partial\over\partial x_i},
\nonumber\\
\left\{q_{2\dot a},q_{2\dot b}\right\}
\teq{-i\over 2\dot x_-^2}\delta_{\dot a\dot b}
\oint ds\,\left(\dot x_i{\partial\over\partial x_i}
+\dot x_+{\partial\over\partial x_+}
-\zeta{d\over ds}{\partial\over\partial\zeta}\right)
\nonumber\\
\teq{i\over 2\dot x_-}\delta_{\dot a\dot b}
\oint ds\,{\partial\over\partial x_-},
\eer
where the last equality can be easily proven, and is related to 
reparametrization invariance. Those commutators agree with those 
of the supercharges up to a sign, as they should.

Note that $q_1$ is not the same as $\delta/\delta \zeta$ 
(\ref{fcvar}), rather it has the opposite sign on the second term. 
This is similar to $\cN=1$ supersymmetry in four dimensions where the 
SUSY generator is the same as the fermionic covariant derivative 
up to such a sign flip. 

Since the action is invariant under supersymmetry, one might have 
hoped to use the supersymmetry generators to write loop equations. 
But the supercharges cannot be expressed in terms of constrained 
variations, therefore this is impossible. Instead the constrained 
variations are
\ber
\oint ds {\delta\over\delta\zeta}
\teq 
q_1+\oint ds {i\zeta\over2\dot x_-}{\partial\over\partial x_+}
=Q_1-\widetilde Q_1,
\\
\oint ds\dot x_i\gamma_i{\delta\over\delta\zeta}
\teq
q_2+\oint ds{i\gamma_i\zeta\over 2\sqrt2\dot x_-}
{\partial\over\partial x_i}
=Q_2-\widetilde Q_2,
\eer
but since those are not symmetries of the action, they don't annihilate 
the loop.

One can also define 
\ber
q_3\teq\sqrt2\dot x_-\oint ds\,\zeta,
\\
q_4\teq\oint ds\,\dot x_i\gamma_i\zeta.
\eer
Among themselves they commute to zero, but with $q_1$ and $q_2$
\ber
\left\{q_{1a},q_{1b}\right\}\teq\sqrt2\dot x_-\delta_{ab},
\nonumber\\
\left\{q_{1a},q_{4\dot b}\right\}
\teq\left\{q_{2\dot b},q_{3a}\right\}=\dot x_i\gamma_{ia\dot b},
\nonumber\\
\left\{q_{2\dot a},q_{4\dot b}\right\}
\teq-\sqrt2\dot x_+\delta_{\dot a\dot b},
\eer
and the right hand side is the light-cone decomposition of 
$\Gamma^\mu\dot x_\mu$.

\subsection{Fermionic Loop Equation}

Another very interesting fermionic operator is
\beq
\hat D =
\lim_{\eta\to0}\oint ds\int_{s-\eta}^{s+\eta}ds'\,
\gamma_i{\delta\over\delta x_i(s)}{\delta\over\delta\zeta(s')}.
\eeq
As we show in the appendix, $\hat D$ acting on the Wilson loop gives a 
fermionic loop equation.  The $\zeta=0$ piece inserts into the loop
\beq
\inv{2^{5\over4}\sqrt{\dot x_-}}\dot x^\mu\Gamma_\mu\Gamma_\nu D^\nu\Psi.
\label{projected-eqn}
\eeq
Again we use the Schwinger-Dyson equations related to the fermionic 
equations of motion
\beq
{\delta\cS\over\delta\bar\Psi}
=-{i\over g_{ym}^2}\Gamma_\mu D^\mu\Psi.
\eeq
The result is two terms proportional to $\dot x_i\gamma_i\zeta$ 
with a relative minus sign.  So if there is no self intersection it's 
zero.

The piece linear in $\zeta$ in $\hat D$ gives
\beq
-\quart\left(D^\mu F_{i\mu}
-\half\left\{\bar\Psi,\Gamma_i\Psi\right\}
-{\dot x_i\over\dot x_-}\left(D^\mu F_{-\mu}
-\half
\left\{\bar\Psi,\Gamma_-\Psi\right\}\right)\right)
\gamma_i\zeta.
\eeq
This is a linear combination of the bosonic equations of motion. 
It is nice to see a different linear combination than
$\dot x^\nu (D^\mu F_{\nu\mu}-\half\{\bar\Psi,\Gamma_\nu\Psi\})$ 
which we got for the supersymmetrized loop Laplacian.
Again if you continue to the next step you find that each term gives 
$\dot x_i\gamma_i\zeta$, and again at smooth points they are 
canceled by the relative minus sign.

One might have actually expected two loop equations based on the 
fer\-mionic equation of motion, but we can get only one. Since 
$\dot x^2=0$ the factor $\dot x^\mu\Gamma^\mu$ in 
(\ref{projected-eqn}) projects onto half the equations. We did 
get a different fermionic loop equation in the 
piece linear in $\zeta$ in $\hat L^{lc}$ (\ref{hat-L1}).

Since $q_1$ and $q_2$ are symmetries of the loop, anticommuting them 
with $\hat D$ should give an operator which annihilates the loop. 
The anticommutator with $q_1$ is trivial, but with $q_2$ one 
gets (after some algebra) the light-cone loop Laplacian
\ber
\left\{q_{1a},\hat D_{\dot b}\right\}\teq 0,
\nonumber\\
\left\{q_{2\dot a},\hat D_{\dot b}\right\}
\teq
{-i\over2\sqrt2\dot x_-}\delta_{\dot a\dot b}\hat L^{lc}.
\eer

\mysection{Conclusions}

Our main result is the operator $\hat L^{lc}$ defined in 
(\ref{hat-L}), which commutes with the constraint 
$\dot x_\mu\dot x^\mu+{i\over2}\zeta\dot\zeta=0$ and when acting 
on a Wilson loop gives a closed loop equation. When the loop is 
smooth the equation reads
\beq
\hat L^{lc}\vev{W}
=-i{\lambda\over\epsilon^3}\oint ds
\left(\dot x^\mu\dot x_\mu +{i\over2}\zeta\dot\zeta\right)
\vev{W}
\sim 0.
\label{equation}
\eeq
Where $\sim$ means that the most divergent pieces cancel. With cusps 
or intersections there will be more divergent terms proportional to 
the product of the two tangent vectors.

There is also a fermionic operator $\hat D$, a ``Dirac operator'' on 
loop space, which also annihilates the loop. The operator 
$\hat L^{lc}$ is the anticommutator of $\hat D$ with a supersymmetry 
generator.

The $AdS$/CFT correspondence tells us how to calculate some Wilson 
loops at strong 't Hooft coupling using minimal surfaces in $AdS$. 
The ansatz is
\beq
\vev{W}\sim\exp\left(-\sqrt\lambda\tilde A\right),
\label{ads}
\eeq
with $\tilde A$ a Legendre transform of the area of the minimal 
surface ending along the loop at the boundary of $AdS$.

For smooth loops 
this calculation, in terms of a minimal surface, is valid only for 
loops satisfying the constraint $\dot x_\mu\dot x^\mu=0$ (where 
$\dot x^\mu$ includes the couplings to both scalars and gauge 
fields). Therefore we can check if the ansatz solves the light-cone 
loop equation.

The calculation is very simple. To leading order in $\lambda$ 
we get
\beq
\hat L^{lc}\vev{W}\sim\lambda\int ds
\left({\delta\tilde A\over\delta x^i(s)}\right)^2\vev{W}.
\eeq
It was argued \cite{Drukker:1999zq} that, as long as the 
constraint is satisfied, the expectation value of the loop is 
not divergent. Since $\delta/\delta x^i$ commutes with the 
constraint, it will give a finite result when acting on 
$\tilde A$. That means that there are no divergent terms, as 
in (\ref{equation}).

This is not a very strong test of the conjecture, since those 
equations are satisfied by any functional that assigns those 
loops a finite value. But it does agree with the assertion that 
those loops are finite, and that the leading behavior is 
proportional to $\sqrt\lambda$.

It would be more interesting to check the equation for loops with 
cusps and intersections, where the right hand side is not zero. 
Unfortunately, those loops can be evaluated by minimal surfaces 
only when $\dot x^2\neq0$. So the new loop equations are not valid. 
There were problems with using the other form of the loop equations 
in this case too.

The calculation with the regular loop Laplacian required extending 
the ansatz (\ref{ads}) to loops that break the constraint. Applying 
the light-cone loop Laplacian does not require this extension, and 
is therefore more reliable.

This argument applies to the $AdS$ calculation with both Lorentzian 
and Euclidean signature. In the Euclidean theory there are still 
light-like lines, since the parameters that couple to the scalars 
are imaginary.

We derived the light-cone loop equations for bosonic theories, where 
they close only for straight light-like lines. There might be 
applications of those loop equations for QCD. For example, it might be 
possible to apply them to Wilson loops wrapping the compact light-like 
direction in DLCQ of gauge theories. For the $\cN=4$ theory they 
close for any loop satisfying the constraint. We did not check if and 
when one can write such loop equations for theories with $\cN=1$ or 
$\cN=2$ supersymmetry, which could be interesting.

When loop equations were first constructed there was a hope to use them 
to quantize the QCD string. This program was never completed, yet it 
is tempting to think that if the regular loop equations are related to 
covariant quantization of the QCD string, the light-cone equations are 
related to light-cone quantization of the string. Since light-cone 
quantization is easier in some cases, it might enable progress in this 
direction. In the case of $\cN=4$ SYM, by the Maldacena conjecture, 
the QCD string is just a fundamental string in $AdS$ background.   

It would be interesting to generalize to arbitrary loops. That would 
require an operator which commutes with a constraint like 
$\dot x^2=m^2$ and gives a closed loop equation.

Also, the equations were checked for the few lowest orders in the 
fermionic parameter $\zeta$. It would be nice to find a formulation 
that will prove the equations to all orders in $\zeta$.

\section*{Acknowledgments}
We would like to thank Hirosi Ooguri for collaboration in the initial 
stages of this work, and David Gross for a lot of help and guidance. 
We also thank the Racah Institute
and the Institute for Advanced Studies at the Hebrew University for 
their hospitality during the course of this work. This work is 
supported in part by the NSF under grant PHY94-07194.

\appendix

\mysection{Filling in Some Details}
\subsection{Conventions}

We use signature $(-,+,\cdots,+)$ and light cone coordinates
\beq
\dot x^\pm={1\over\sqrt2}\left(\pm\dot x^0+\dot x^1\right).
\eeq

The Lagrangian is
\beq
\cL=-{1\over g_{YM}^2}\Tr\left(\quart F_{\mu\nu}F^{\mu\nu}
+{i\over 2}\bar\Psi\Gamma^\mu D_\mu\Psi\right),
\eeq
$D_\mu=\partial_\mu-iA_\mu$ and
$F_{\mu\nu}=\partial_\mu A_\nu-\partial_\nu A_\mu-i[A_\mu,\,A_\nu]$.

This Lagrangian is invariant under the supersymmetry transformation
\ber
\delta A_\mu \teq {i\over 2}\bar\epsilon\Gamma_\mu\Psi,
\nonumber\\
\delta\Psi \teq -\quart F_{\mu\nu}\Gamma^{\mu\nu}\epsilon.
\eer
The equations of motion are
\ber
{\delta\cS\over\delta A^\mu}
\teq
-\inv{g_{ym}^2}\left(D^\nu F_{\mu\nu}
-{1\over 2}\left\{\bar\Psi,\,\Gamma_\mu\Psi\right\}\right)
=0,
\nonumber\\
{\delta\cS\over\delta\bar\Psi}
\teq
-{i\over g_{ym}^2}\Gamma^\mu D_\mu\Psi
=0.
\eer

We decompose $\Psi$ in the light-cone gauge into two spinors
$\Psi_1$ and $\Psi_2$ in the $8_s$ and $8_c$ of $SO(8)$ respectively.
Under this decomposition
\beq
\matrix{
\Gamma^0\Gamma_i=\pmatrix{0&\gamma_i\cr \gamma_i^T&0}&
\Gamma^0\Gamma_+=\sqrt{2}\pmatrix{1&0\cr 0&0}&
\Gamma^0\Gamma_-=\sqrt{2}\pmatrix{0&0\cr 0&-1}\cr
\Gamma_{ij}=\pmatrix{\gamma_{ij}&0\cr 0&\gamma_{ij}}&
\Gamma_{+i}=\sqrt{2}\pmatrix{0&0\cr -\gamma_i^T&0}&
\Gamma_{-i}=\sqrt{2}\pmatrix{0&\gamma_i\cr 0&0}\cr
\Gamma_{+-}=\pmatrix{-1&0\cr 0&1}&
\Gamma_{-+}=\pmatrix{1&0\cr 0&-1},&
}
\eeq
where $\gamma_i$ are the Clebsch-Gordan coefficients of $SO(8)$.

For example
\beq
\bar\Psi\Gamma^\mu D_\mu\Psi
=\sqrt{2}\Psi_1 D_-\Psi_1
-\sqrt{2}\Psi_2 D_+\Psi_2
+\Psi_1\gamma_iD_i\Psi_2
+\Psi_2\gamma_i^TD_i\Psi_1.
\eeq

Let us write again what appears in the exponent of the supersymmetric 
Wilson loop (\ref{exponent})
\ber
i\cA_\mu\dot x^\mu
\teq
iA_\mu \dot x^\mu
-{\sqrt{\dot x_-}\over 2^{5\over 4}}
\left(\sqrt{2}\zeta\Psi_1
+{\dot x_i\over\dot x_-}\zeta\gamma_i\Psi_2\right)
+{3\over 16\dot x_-}
\dot x_{[i} F_{j-]}\zeta\gamma_{ij}\zeta
\nonumber\\
\tab
+{i\over 2^\quart48\sqrt{\dot x_-}}
\zeta\gamma_{ij}\zeta
\left(D_i-{\dot x_i\over\dot x_-}D_-\right)\zeta\gamma_j\Psi_2
+\ldots
\eer
and $\zeta\dot\zeta$ was absorbed into $\dot x_+$. Then the 
loops we are studying satisfy the constraint
$2\dot x_+\dot x_-+\dot x_i^2+{i\over2}\zeta\dot\zeta=0$. 
To express that we use the functional variations which 
commute with the constraint
\beq
{\delta\over\delta x_i}
\sim
{\partial\over\partial x_i}
-{\dot x_i\over\dot x_-}{\partial\over\partial x_+}
\qquad\hbox{and}\qquad
{\delta\over\delta \zeta}
\sim
{\partial\over\partial \zeta}
+{i\zeta\over4\dot x_-}{\partial\over\partial x_+},
\eeq
where the derivatives on the right hand side are unconstrained 
functional variations. As was explained in the text, there are 
corrections to this equation for higher derivatives.

\subsection{$q_1$}
The first identity is for the supersymmetry generator $Q_1$ defined 
in (\ref{Q1})
\beq
Q_1
=q_1
={\partial\over\partial\zeta}
-{i\zeta\over4\dot x_-}{\partial\over\partial x_+}.
\eeq
To demonstrate this relation we calculate the first few terms 
in an expansion in $\zeta$
\ber
\left.Q_1\!\right._{[0]}
\teq
-{1\over 2^{5\over 4}\sqrt{\dot x_-}}
\left(\sqrt{2}\dot x_-\Psi_1
+\dot x_i\gamma_i\Psi_2\right),
\nonumber\\
\left.Q_1\!\right._{[1]}
\teq
-{1\over 4\dot x_-}
\left(\dot x_i F_{j-}
+\half\dot x_- F_{ij}\right)
\zeta\gamma_{ij}
-{1\over 4\dot x_-}\dot x^\mu F_{\mu -}\zeta
\nonumber\\
\teq
{3\over 8\dot x_-}
\dot x_{[i} F_{j-]}\gamma_{ij}\zeta
-{1\over 4\dot x_-}\dot x^\mu F_{\mu -}\zeta,
\nonumber\\
\left.Q_1\!\right._{[2]}
\teq
{i\over 2^\quart16\sqrt{\dot x_-}}
\left({\dot x_i\over\dot x_-}\gamma_j D_-\Psi_2
+\gamma_i D_j\Psi_2\right)
\zeta\gamma_{ij}\zeta
\nonumber\\
\teq
{i\over 2^\quart16\sqrt{\dot x_-}}
\zeta\gamma_{ij}\zeta
\left(D_i-{\dot x_i\over\dot x_-}D_-\right)\gamma_j\Psi_2.
\eer
On the other hand
\ber
\left.q_1\!\right._{[0]}
\teq
-{1\over 2^{5\over 4}\sqrt{\dot x_-}}\left(
\sqrt{2}\dot x_-\Psi_1
+\dot x_i\gamma_i\Psi_2\right),
\nonumber\\
\left.q_1\!\right._{[1]}
\teq
{3\over 8\dot x_-}
\dot x_{[i} F_{j-]}\gamma_{ij}\zeta
-{1\over4\dot x_-}\dot x^\mu F_{\mu-}\zeta,
\nonumber\\
\left.q_1\!\right._{[2]}
\teq
{i\over 2^\quart48\sqrt{\dot x_-}}
\!
\left(
\zeta\gamma_{ij}\zeta
\left(D_i-{\dot x_i\over\dot x_-}D_-\right)\gamma_j\Psi_2\;
+2\gamma_{ij}\zeta
\left(D_i-{\dot x_i\over\dot x_-}D_-\right)\zeta\gamma_j\Psi_2
\right)
\nonumber\\
\tab
+{i\over2^\quart 8\sqrt{\dot x_-}}\zeta\;D_-
\left(\sqrt{2}\zeta\Psi_1
+{\dot x_i\over\dot x_-}\zeta\gamma_i\Psi_2\right)
\nonumber\\
\teq
{i\over 2^\quart16\sqrt{\dot x_-}}
\zeta\gamma_{ij}\zeta
\left(D_i-{\dot x_i\over\dot x_-}D_-\right)\gamma_j\Psi_2\;
\nonumber\\
\tab
+{i\over 2^\quart8\sqrt{\dot x_-}}
\zeta
\left[
\left(D_i-{\dot x_i\over\dot x_-}D_-\right)\zeta\gamma_i\Psi_2\;
+D_-\left(\sqrt{2}\zeta\Psi_1
+{\dot x_i\over\dot x_-}\zeta\gamma_i\Psi_2\right)
\right],
\nonumber\\
\label{q1-expand}
\eer
where we used the Fierz identity 
$\gamma_{ia\dot a}\gamma_{ib\dot b}+\gamma_{ia\dot b}\gamma_{ib\dot a} 
=2\delta_{ab}\delta_{\dot a\dot b}$, which is related by triality to 
the usual Clifford algebra relation 
$\gamma_i\gamma_j^T+\gamma_j\gamma_i^T=2\delta_{ij}$
\ber
\gamma_{ij}\zeta\;\zeta\gamma_j\Psi_2
=
-\zeta\gamma_i\gamma_j^T\;\zeta\gamma_j\Psi_2
+\zeta\;\zeta\gamma_i\Psi_2
=\zeta\gamma_{ij}\zeta\;\gamma_j\Psi_2
+3\zeta\;\zeta\gamma_i\Psi_2.
\eer
The last term on the right hand side of (\ref{q1-expand}) is zero (for 
a smooth loop) by the fermionic loop equation
\ber
\sqrt{2}\zeta D_-\Psi_1
+\zeta\gamma_i D_i\Psi_2
=\zeta\left(\Gamma^0\Gamma^\mu D_\mu\Psi\right)_1
=\zeta{\delta\over\delta\Psi_1}
\sim \zeta\zeta
=0.
\eer
We see therefore that $Q_1=q_1$.

\subsection{$q_2$}

Next we show
\beq
Q_2
=
q_2
=
{1\over\sqrt2\dot x_-}
\left(\dot x_i\gamma_i q_1
-{i\over 2}{\delta\over\delta x_i}\gamma_i\zeta\right).
\eeq
We act with $Q_2$ (\ref{Q2}) on the loop and expand
\ber
\left.Q_2\!\right._{[0]}
\teq
{1\over2^{5\over 4}\sqrt{\dot x_-}}\left(
\sqrt{2}\dot x_+\Psi_2
-\dot x_i\gamma_i\Psi_1
\right),
\nonumber\\
\left.Q_2\!\right._{[1]}
\teq
-{1\over 4\sqrt{2}\dot x_-}\left(
2\dot x_-F_{+i}
-\dot x_jF_{ji}
+\dot x_iF_{-+}
\right)\gamma_i\zeta
-{i\over 8\sqrt{2}\dot x_-}
\dot x_iF_{jk}\zeta\gamma_{ijk}
\nonumber\\
\teq
-{1\over 4\sqrt{2}\dot x_-}
\left(3\dot x_{[-}F_{+i]}
+\dot x^\mu F_{\mu i}
\right)\gamma_i\zeta
-{1\over 8\sqrt{2}\dot x_-}
\dot x_iF_{jk}\zeta\gamma_{ijk}.
\eer
On the other hand
\ber
{1\over\sqrt2}{\dot x_i\over\dot x_-}
\gamma_i\left.q_1\!\right._{[0]}
\teq
-{1\over 2^{3\over 4}\dot x_-^{3\over2}}
\left(\sqrt{2}\dot x_i\gamma_i\dot x_-\Psi_1
+\dot x_i^2\Psi_2\right)
\nonumber\\
\teq
{1\over 2^{5\over 4}\sqrt{\dot x_-}}
\left(\sqrt{2}\dot x_+\Psi_2
-\dot x_i\gamma_i\Psi_1\right),
\nonumber\\
{1\over\sqrt2}{\dot x_i\over\dot x_-}
\gamma_i\left.q_1\!\right._{[1]}
\teq
{1\over 8\sqrt2\dot x_-^2}
\left(3\dot x_k\dot x_{[i} F_{j-]}\gamma_k\gamma_{ij}\zeta
-2\dot x_i\dot x^\mu F_{\mu -}\gamma_i\zeta\right)
\nonumber\\
\teq
-{1\over 4\sqrt2\dot x_-}
\left(3\dot x_{[i}F_{-+]}
-\dot x^\mu F_{\mu i}
+2{\dot x_i\dot x^\mu\over\dot x_-}F_{\mu-}\right)\gamma_i\zeta
\nonumber\\
\tab
+{1\over 8\sqrt2\dot x_-}\dot x_i F_{jk}\gamma_{ijk}\zeta,
\eer
and
\beq
i{\delta\over\delta x_i}\gamma_i\zeta_{[1]}
=
\dot x^\mu F_{\mu i}\gamma_i\zeta
-{\dot x^\mu\dot x_i\over\dot x_-}F_{\mu -}\gamma_i\zeta.
\eeq

Together
\ber
\left.q_2\!\right._{[0]}
\teq
{1\over 2^{5\over 4}\sqrt{\dot x_-}}
\left(\sqrt{2}\dot x_+\Psi_2
-\dot x_i\gamma_i\Psi_1\right),
\nonumber\\
\left.q_2\!\right._{[1]}
\teq
{1\over 8\sqrt2}{\dot x_i\over\dot x_-} F_{jk}\gamma_{ijk}\zeta
-{1\over 4\sqrt2\dot x_-}
\left(3\dot x_{[i}F_{-+]}
+\dot x^\mu F_{\mu i}\right)\gamma_i\zeta.
\phantom{(A.21)}
\eer
So $Q_2=q_2$.

\subsection{$\hat D$}

Finally we calculate
\beq
\hat D=\gamma_i{\delta\over\delta x^i}{\delta\over\delta\zeta},
\eeq
which gives the fermionic loop equation.
\ber
\hat D_{[0]}
\teq
\gamma_i
\left({\partial\over\partial x^i}
-{\dot x_i\over\dot x_-}{\partial\over\partial x_+}\right)
{\partial\over\partial\zeta}_{[0]}
\nonumber\\
\teq
-{1\over 2^{5\over 4}\sqrt{\dot x_-}}
\left(D_i-{\dot x_i\over\dot x_-}D_-\right)
\gamma_i
\left(\sqrt{2}\dot x_-\Psi_1
+\dot x_j\gamma_j\Psi_2\right)
\nonumber\\
\teq
-{1\over 2^{5\over 4}\sqrt{\dot x_-}}\left(
\sqrt{2}\dot x_-\gamma_iD_i\Psi_1
-\sqrt{2}\dot x_i\gamma_iD_-\Psi_1
+\dot x_j\gamma_i\gamma_jD_i\Psi_2
\phantom{\dot x_i^2\over\dot x_-}\right.\nonumber\\
\tab\left.
-8 {d\over ds}\Psi_2
-{\dot x_i^2\over\dot x_-}D_-\Psi_2
\right)
\nonumber\\
\teq
{1\over 2^{5\over 4}\sqrt{\dot x_-}}\left(
\sqrt{2}\gamma_i\left(\dot x_iD_--\dot x_-D_i\right)\Psi_1
+\left(\dot x_j\gamma_j\gamma_i D_i
+2\dot x_-D_+
\right)\Psi_2
\right)
\nonumber\\
\tab
-{1\over2^\quart\sqrt{\dot x_-}}\dot x^\mu D_\mu\Psi_2
\nonumber\\
\teq
{\dot x^\mu\over 2^{5\over 4}\sqrt{\dot x_-}}
\left(\Gamma_\mu\Gamma_\nu D^\nu\Psi\right)_2
-{1\over2^\quart\sqrt{\dot x_-}}\dot x^\mu D_\mu\Psi_2.
\eer
We dropped a total derivative in the course of the calculation. 
The first term gives a fermionic loop equation
\beq
\left(\dot x^\mu\Gamma_\mu\Gamma_\nu D^\nu\Psi\right)_2
=\left(\dot x^\mu\Gamma_\mu{\delta\over\delta\bar\Psi}\right)_2
\propto \dot x^\mu\dot x^\nu\Gamma_\mu\Gamma_\nu
=0,
\eeq
plus nonzero terms at cusps and intersections. 
The second term is almost 
a total derivative, $\dot x^\mu D_\mu$ is the bosonic part of 
$d/ds$. Since we will look at the term linear in 
$\zeta$ next, we will need
\beq
\dot x^\mu D_\mu\Psi_2
=
{d\over ds}\Psi_2
-{\sqrt{\dot x_-}\over2^{5\over4}}
\left(\sqrt{2}\zeta\Psi_1
+{\dot x_i\over\dot x_-}\zeta\gamma_i\Psi_2\right)\Psi_2
+\ldots
\eeq

The term linear in $\zeta$ contains
\ber
\tab\hskip-.2in
\gamma_i
\left({\partial\over\partial x^i}
-{\dot x_i\over\dot x_-}{\partial\over\partial x_+}\right)_{[0]}
\left({\partial\over\partial\zeta}
+{i\zeta\over4\dot x_-}{\partial\over\partial x_+}\right)_{[1]}
\nonumber\\
\teq
{1\over 8\dot x_-}\left(
2\dot x_j D_iF_{k-}
-2{\dot x_i\dot x_j\over\dot x_-} D_-F_{k-}
+\dot x_- D_iF_{jk}
-\dot x_i D_-F_{jk}
\right)
\gamma_i\gamma_{jk}\zeta
\nonumber\\
\tab
+{1\over4\dot x_-}
\left(\dot x^\mu D_iF_{\mu-}
-{\dot x^\mu\dot x_i\over\dot x_-}D_-F_{\mu-}\right)\gamma_i\zeta
-{2\over\dot x_-}{d\over ds}F_{i-}\gamma_i\zeta
\nonumber\\
\teq
-\quart
\left(D^\mu F_{i\mu}
-{\dot x_i\over\dot x_-}D^\mu F_{-\mu}\right)\gamma_i\zeta.
\label{ddeta}
\eer
There are two more contributions at this order with two $\Psi$'s 
and a $\zeta$. One was mentioned above, the other comes from 
the fermionic part of $\delta/\delta x^i$
\ber
\tab\hskip-.2in
\gamma_i
\left({\partial\over\partial x^i}
-{\dot x_i\over\dot x_-}{\partial\over\partial x_+}\right)_{[1]}
{\partial\over\partial\zeta}_{[0]}
+{1\over2\sqrt2}\left(\sqrt{2}\zeta\Psi_1
+{\dot x_i\over\dot x_-}\zeta\gamma_i\Psi_2\right)\Psi_2
\nonumber\\
\teq
-{1\over 4\sqrt{2}}
\left(\left(\zeta\gamma_i\Psi_2\right)
\gamma_i
\left(\sqrt{2}\Psi_1
+{\dot x_j\over\dot x_-}\gamma_j\Psi_2\right)
-2\left(\sqrt{2}\zeta\Psi_1
+{\dot x_i\over\dot x_-}\zeta\gamma_i\Psi_2\right)\Psi_2\right)
\nonumber\\
\teq
{1\over 4\sqrt{2}}
\gamma_i\zeta
\left(
\sqrt{2}\Psi_2\gamma_i\Psi_1
+{\dot x_i\over\dot x_-}\Psi_2\Psi_2\right),
\label{psipsieta}
\eer
by a Fierz identity. In those equations an anticommutator of the 
$\Psi$'s is implied, which eliminates terms like 
$\Psi_2\gamma_{ij}\Psi_2$.

Together (\ref{ddeta}) and (\ref{psipsieta}) give
\beq
\gamma_i{\delta\over\delta x^i}{\delta\over\delta\zeta}_{[1]}
=
-{\gamma_i\zeta\over4}
\left(
D^\mu F_{i\mu}
-\Psi_2\gamma_i\Psi_1
-{\dot x_i\over\dot x_-}D^\mu F_{-\mu}
-{1\over\sqrt{2}}{\dot x_i\over\dot x_-}\Psi_2\Psi_2\right).
\eeq
This is a linear combination of the bosonic equations of motion 
which vanishes, by a loop equation, at a smooth point.

\end{document}